\documentclass[useAMS,usenatbib]{mn2e}
\usepackage{natbib}
\usepackage{float}
\usepackage{graphicx}
\usepackage{epstopdf}
\usepackage{amsmath}
\usepackage{placeins}
\newcommand{\los}{line-of-sight}

\title[Dust within the nuclear star cluster in the Milky Way]{Dust within the nuclear
 star cluster in the Milky Way} \author[S. Chatzopoulos et al.]{S. Chatzopoulos$^{1}$\thanks{E-mail: sotiris@mpe.mpg.de,
 gerhard@mpe.mpg.de}, O. Gerhard$^{1}$, T. K. Fritz$^{2}$, C. Wegg$^{1}$, S. Gillessen$^{1}$, O. Pfuhl$^{1}$,\and
 F. Eisenhauer$^{1}$
  \\
  $^{1}$Max Planck Institut fur Extraterrestrische Physic, Postfach
  1312, D-85741, Garching, Germany
  \\
  $^{2}$Department of Astronomy, University of Virginia, 530 McCormick Road, Charlottesville VA 22904-4325, USA}

\begin{document}

\date{Submitted 2015 April 14}

\pagerange{\pageref{firstpage}--\pageref{lastpage}} \pubyear{2015}

\maketitle

\label{firstpage}

\begin{abstract}
  
The mean absolute extinction towards the central parsec of the Milky
Way is $A_{K}\simeq 3$ mag, including both foreground and Galactic center
dust. Here we present a measurement of dust extinction \emph{within} the
Galactic old nuclear star cluster (NSC), based on combining differential
extinctions of NSC stars with their $\upsilon_l$ proper motions along Galactic longitude.
Extinction within the NSC preferentially affects stars at its far side,
and because the NSC rotates, this causes higher extinctions for NSC stars with
negative $\upsilon_l$, as well as an asymmetry in the $\upsilon_l$-histograms.
We model these effects using an axisymmetric dynamical model of the NSC in
combination with simple models for the dust distribution. Comparing
the predicted asymmetry to data for $\sim7'100$ stars in several NSC fields,
we find that dust associated with the Galactic center mini-spiral with extinction
$A_K\simeq 0.15-0.8$ mag explains most of the data. The largest extinction
$A_K\simeq0.8$ mag is found in the region of the Western arm of the mini-spiral.
Comparing with total $A_K$ determined from stellar colors, we determine the extinction
in front of the NSC. Finally, we estimate that for a typical extinction of $A_K\simeq0.4$
the statistical parallax of the NSC changes by $\sim 0.4\%$.
  
\end{abstract}


\section{INTRODUCTION}

Nuclear star clusters (NSC) are located at the centers of most spiral galaxies \citep{c1998,
  boeker2002}.  Their study became possible via high spatial resolution observations from HST in the
1990s. They have properties similar to those of globular clusters although they are more compact,
more massive and on average 4 mag brighter than the old globular clusters of the Milky Way
\citep{boeker2004,walcher2005}. Many NSCs host an AGN \citep{seth2008} i.e. a supermassive black
hole (SMBH) in their centers, have complex star formation histories \citep{rossa2006,seth2006} and
obey scaling-relations with host galaxy properties as do central SMBHs
\citep{ferrarese2006,wehner2006}.

The study of NSCs is of great interest because several of the most extreme physical phenomena occur
within them such as SMBHs, active galactic nuclei, star-bursts and extreme stellar densities.  The
Galactic NSC is particularly interesting because of its proximity. At a distance of about 8kpc from
Earth it is the only NSC in which individual stars can be resolved.

The center of the galactic NSC harbors a SMBH \citep{geisen2010, ghez2008}. Joint statistical
analysis based on orbits around Sgr A* \citep{ge2009}, star counts and kinematic data gives
${M_\bullet} = (4.23\pm0.14)\times {10^6}{M_\odot}$ and a statistical parallax ${R_0} \!=\! 8.33
\!\pm\! 0.11$ kpc \citep{cf2015}. Recent studies \citep{schodel2014, cf2015} have revealed that the
NSC is flattened with an axial ratio $q\approx0.73$, which is consistent with the kinematic data
\citep{cf2015}.

Observations of the NSC at optical-UV wavelengths are not feasible because of the high extinction
$A_V\ge30$ mag due to interstellar dust \citep{ss2003,fc2011}. Therefore we rely on the infrared,
with average K-band extinction toward the central parsec close to $A_K \approx 3$
\citep{rr1988,schoedel2010,fc2011}.  This is mostly foreground extinction; however, it is very
difficult observationally to measure the extinction variation along the line-of-sight within the
NSC.

The area around Sgr A* contains ionized gas which can be well described by a system of ionized
streamers or filaments orbiting Sgr A* \citep{e1983,sl1985} that is also associated with hot
  dust \citep{lau2013}. This complex structure of ionized gas is called the 'mini-spiral' and
consists of four main components: the northern arm, the eastern arm, the western arm and the bar
\citep{zm2009}, surrounded by the circumnuclear disk of inner radius $\sim1.6$pc
\citep{jg1993,cs2005}.

In our previous dynamical study of the NSC, an asymmetry in the $\upsilon_l$-proper motions was
observed in the histograms which was attributed to dust causing stars on the far side of the NSC to
fall out of the sample.  The aim of this paper is to present estimates for the extinction within the
NSC based on our dynamical model and see to what extent this is correlated with the mini-spiral, to
try to understand the slight asymmetry in the $\upsilon_l$ velocity histograms of the NSC, and to
check the impact of this on the fundamental parameters derived from the dynamical model such as the
mass and the distance.

In section 2 we discuss briefly our recent dynamical model of the NSC and describe
qualitatively the effects of dust on the observed dynamics of the NSC. In section 3 we show evidence
based on the dynamics for the presence of dust within the NSC. In section 4 we develop a method for
making an analytical model for the dust extinction that can be used on top of an existing dynamical
model. Finally in section 5 we present extinction values for the dust within the NSC based on the
prediction of the model in conjunction with the mini-spiral observations.

\begin{figure*}
\centering
\includegraphics[width=0.8\textwidth]{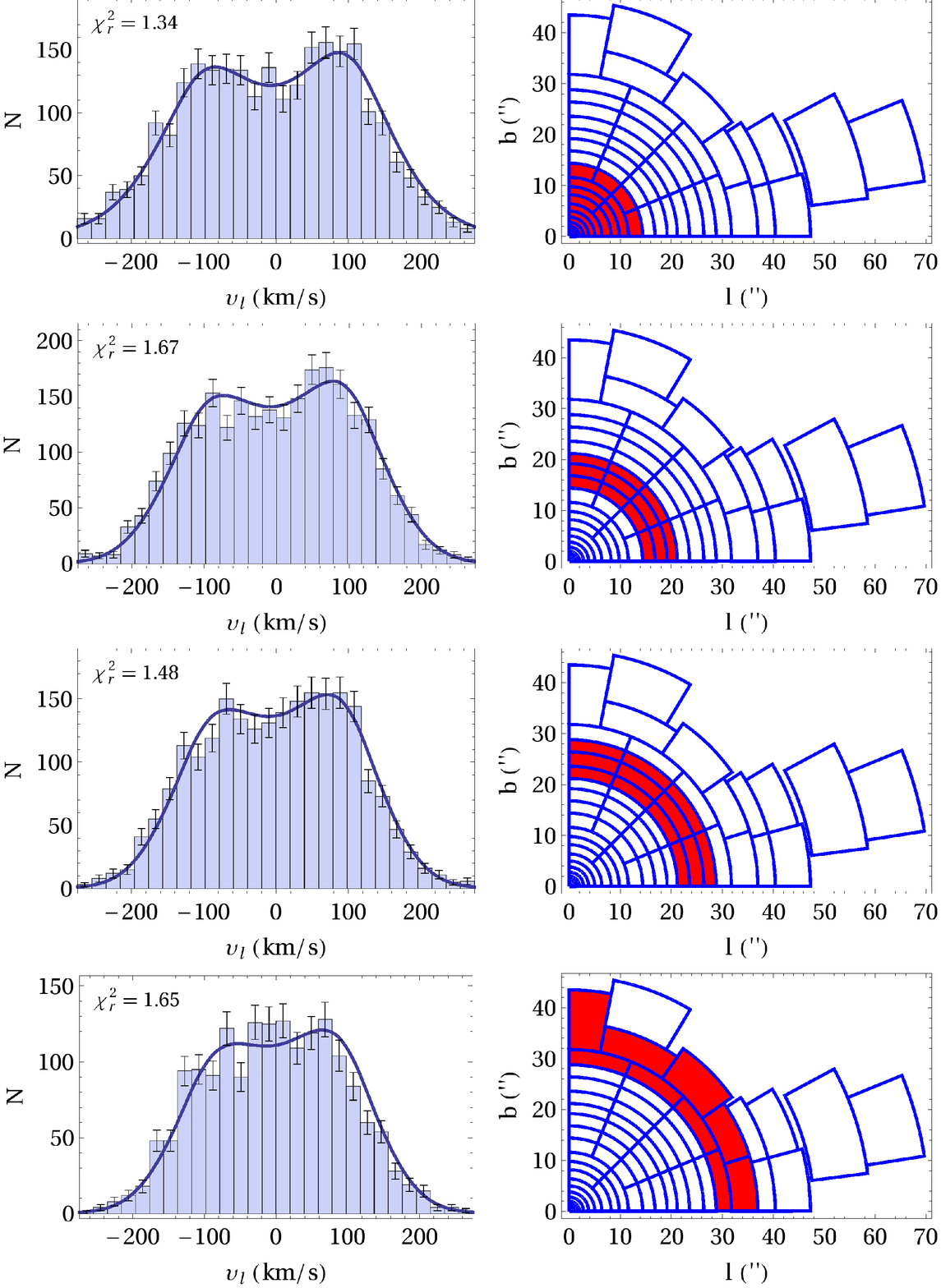}
\caption{Histograms of $\upsilon_l$ proper motions and VPs predicted by an axisymmetric dynamical
model of the NSC combined with a dust extinction model. The data and cells are as in \citet{cf2015}. Each
star is mapped to the first quadrant using $(l,b)\to(|l|,|b|)$. The smooth blue lines are derived
from the recent model of \citet{cf2015} averaged over the cells combined as shown in the right
panels, to which is added a homogeneous dust model with total $A_{K} =0.4$, extending from
  $-100''$ to $+100''$ along the \los (see Section 5).}
\label{plot7}
\end{figure*}

\section[]{Effects of dust on the apparent dynamics of the NSC}

In this section we give a brief description of the recent dynamical model of the NSC from \cite{cf2015}
(Section 2.1) and we show initial evidence for dust extinction within the NSC (Section 2.2).

\subsection{Axisymmetric dynamical model of the NSC}

The dynamical model of the NSC described in \cite{cf2015} is an axisymmetric model including
flattening and rotation which is an excellent match to histograms of proper motions and
line-of-sight velocities in many bins on the sky.  For the density model we used a two-component
spheroidal $\gamma$-model \citep{d1993,tr1994} which we fitted to the star count surface density
data in ($l$, $b$) provided by \cite{fc2014}.  The inner rounder component can be considered as the
NSC and the outer, more flattened component as the inner part of the nuclear stellar disk. Using the density
we applied axisymmetric Jeans modeling in order to constrain the stellar mass $M_*$, the black hole
mass $M_{_\bullet}$, and the distance $R_0$ of the NSC which we found to be
\begin{align}
\begin{array}{l}
M_*(r<100'')=(8.94 \pm 0.31{|_{\rm stat}}
\pm\!0.9{|_{\rm syst}})\times {10^6}{M_\odot} \\
M_\bullet
=(3.86\!\pm\!0.14{|_{\rm stat} \pm 0.4{|_{\rm syst}}})
\times {10^6}{M_\odot } \\
R_0=8.27 \pm
0.09{|_{\rm stat}}\pm 0.1{|_{\rm syst}} \rm kpc
\end{array}
\end{align}
for the NSC only, not including the constraints from stellar orbits around Srg A*.

Having this information we used the \cite{qh1995} algorithm to calculate the even part of the
2-Integral distribution function (DF) $f(E,L_z)$. This allowed us to calculate the velocity profiles
(VP) of the model. We found that the even part of the DF can predict very well the characteristic
2-peak shape \citep{sm2009, fc2014} of the velocity histograms (VH) for the $\upsilon_l$ proper
motion velocities, as well as the VHs in $\upsilon_b$. The addition of a suitable odd part in $L_z$
to the even part of the DF represents the rotation of the cluster, so that also the $\upsilon_{\rm
 los}$ can be reproduced.

\subsection{Asymmetry of the $\upsilon_l$ proper motion histograms}

Upon a closer look at the velocity histograms in $l$ direction $(\textrm{VH}_l)$ (see
Fig.\ref{plot7}) it is noticeable that the right peak is often slightly higher than the left
peak.  Seemingly there are more stars in the front of the cluster (positive velocities) than in the
back.

The proper motion data are from \cite{fc2014}. They are given in Galactic longitude $l^*$ and
Galactic latitude $b^*$ angles centered on Sgr A*. In the following we always refer to the shifted
coordinates but will omit the asterisks for simplicity. We assume that the rotation axis of the NSC
is aligned with the rotation axis of the Milky Way disk. This is in accordance with the very
symmetric Spitzer surface density distribution of \cite{schodel2014}.

\begin{figure}
\centering
\includegraphics[width=\linewidth]{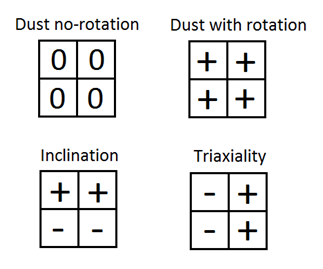}
\caption{ Figure illustrating the velocity profile asymmetries discussed in the text. Each
  small square represents a quadrant in the $(l,b)$ coordinate system, a '$+$' signifies an
  asymmetry in the $\upsilon_l$ VHs (e.g. right peak higher) a '$-$' the opposite asymmetry and
  '$0$' no asymmetry at all. Only dust produces the same asymmetry in every quadrant. }
\label{plot4}
\end{figure}

At least three effects could produce an asymmetry as in Fig.~\ref{plot7}. Figure \ref{plot4}
illustrates the effect of dust extinction without rotation, dust extinction with rotation,
inclination and triaxiality on the VHs. Each small square represents a quadrant of the shifted
Galactic coordinate system $(l,b)$ where Sgr A* is at the center.  A '$+$' signifies an asymmetry
(e.g. right peak higher), a '$-$' the opposite asymmetry, and '$0$' no asymmetry at all (both peaks
same height). We will see in the following Sections that dust with rotation produces the same
asymmetry in every quadrant. Because of the dust fewer stars will be visible at the back of the
cluster, and because of the rotation the missing stars will be stars with negative
velocities. Inclination of the NSC produces opposite results for the upper and lower quadrants,
because when the \los\, does not pass exactly through the center, it passes through areas of unequal
density in front of and behind the NSC. Finally, triaxiality produces opposite results for the right
and left quadrants: consider a triaxial system with long axis in the Galactic plane but rotated
away from the line-of-sight to the observer, with a triaxially symmetric rotation field. If the
stars at the front of the cluster lead to a higher peak for $\upsilon_l>0$ on the $l>0$ side, say,
then by symmetry the opposite will be true on the $l<0$ side.

Therefore, if we symmetrize the data to one quadrant only, the effect of the dust will remain while
the effects of the inclination and triaxiality will cancel out, as illustrated in
Fig.~\ref{plot4}. This is what we observe for the symmetrized data of the NSC
(Fig. \ref{plot7}). Therefore we conclude that the observed asymmetry cannot be a result of
inclination and triaxiality but that it may be produced by dust extinction in conjunction with
rotation. We note here that in order to observe this asymmetry (right peak higher than left) in the
velocity profile in the $l$ direction $(\textrm{VP}_l)$, the dust should be \emph{inside} the
cluster (i.e. within a few parsecs of the Galactic center) where the density is maximum, otherwise
only a change in scale of the VPs would take place.

\section[]{Differential extinction in the NSC}

For the work reported in the following, we use photometric data in the $H$ and $K$ bands and
proper motions for 7101 stars from \cite{fc2014}. We split the data into a central and an
extended field. The central field is a square centered on Sgr A* with size of $40''$ and contains
5847 stars. The rest of the stars belong to the extended fields, as shown in Fig. \ref{plot2}.

\subsection{Total extinction}

We obtain the extinction towards each star from the $H-K$ color (stars without H photometry and early-type stars
are excluded). We obtain  intrinsic color estimates by assuming that the stars are at the distance of the Galactic
Center and that they are giants, as it is the case for most stars in the Galactic Center \citep{pa2014}. The
intrinsic color varies between 0.065 and 0.34 but the majority of stars belong to a small magnitude
range around the red clump. Therefore and also because the extinction is high, the influence of intrinsic color
uncertainties on the extinction is small compared to other effects, like photometric uncertainties. We use the
extinction law of \cite{fc2011}, implying $A_\mathrm{K}=1.348\,E(H-\mathrm{K})$.

In this work we are mainly interested in the extinction variation $A_\mathrm{K}$ \emph{within} the Galactic center.
We obtain an estimate for that by measuring for each star the extinction relative to its neighbors, more
specifically relative to the median extinction of its 15 closest neighbors. Obvious foreground stars were
already excluded in \cite{fc2014}. By using 15 neighbors we obtain a robust median extinction estimate that
is much less affected by extinction variations in the plane of the sky. To further reduce the influence of
this extinction variation  we exclude stars with too few close neighbors.

\begin{figure}
\centering
\includegraphics[width=\linewidth]{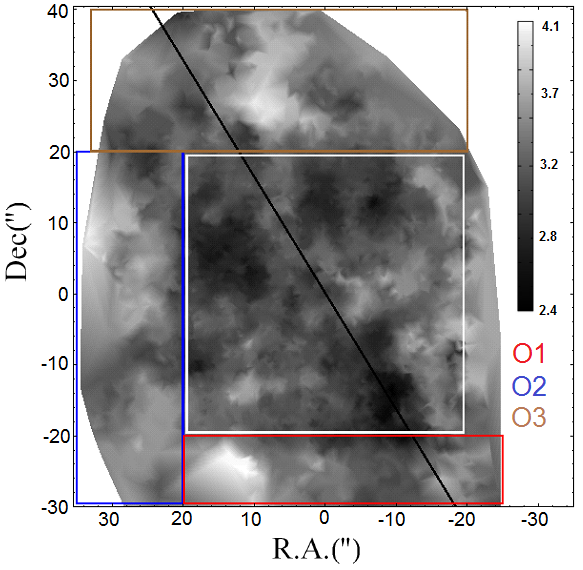}
\caption{Map of $A_K$ for the NSC derived from $H-K$ colors. The map is similar to that of
  \citet{schoedel2010}. The Galactic plane is shown as a black line. The central field is shown in
  white and the outer fields O1-O3 in red, blue and brown.}
\label{plot2}
\end{figure}

\begin{figure}
\centering
\includegraphics[width=\linewidth]{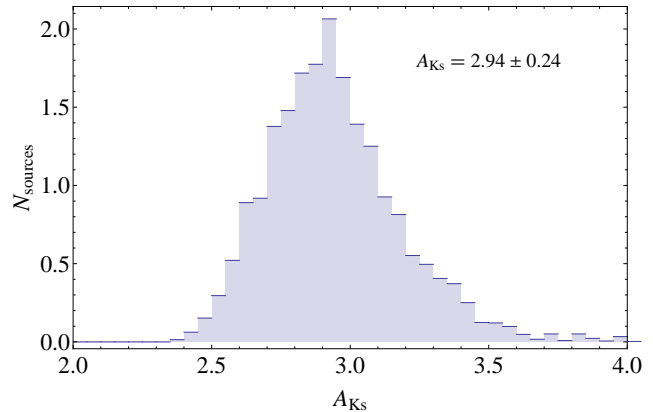}
\caption{Histogram of the extinction $A_{K}$ based on $H-K$ colors for all the stars.
Mean extinction and standard deviation also given. Stars with small $A_K$ are excluded because they are foreground
stars \citep{fc2014}.}
\label{plot1}
\end{figure}

Fig. \ref{plot2} shows a map of interpolated total extinction for the central and extended fields
based on $H-K$ colors. The area within the white frame is the central field which is consistent with
Fig. 6 of \cite{schoedel2010}.  Most of the extinction of typically $A_K=3$ mag shown in this plot
is foreground extinction but a fraction $\simeq0.4$ mag or so is intrinsic to the NSC region as we
show in the following. Fig. \ref{plot1} shows a histogram of the extinction for the central
field. The mean extinction inferred from this plot is $A_K=2.94$ mag with standard deviation $0.24$
mag. This extinction value is very close to the value $A_K=2.74$ measured by \cite{schoedel2010}.
The difference stems partly from the slightly different extinction laws and partly from the difference
in areal coverage. Using the same law as \cite{schoedel2010} results in $A_K=2.85$.

\subsection{Extinction in the NSC region}
\label{sec:NSCextinction}

\begin{figure}
\centering
\includegraphics[width=\linewidth]{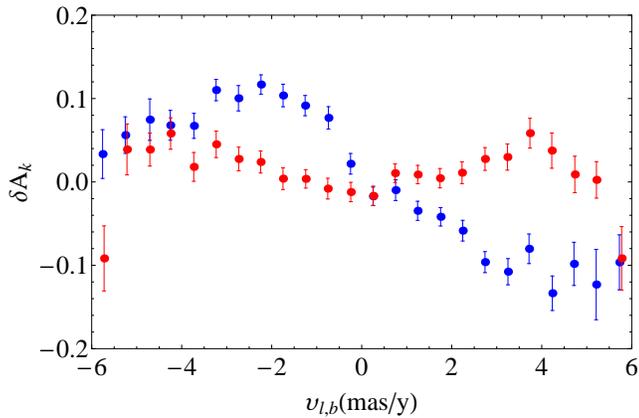}
\caption{Average differential extinction of nuclear cluster stars plotted as a
function of $\upsilon_l$(blue) and $\upsilon_b$(red) proper motion. The differential extinction
is inferred from the difference in the color of a star to the median of its 15 nearest neighbors
using the extinction law of \citet{fc2011} and correcting also for the weak color variation with 
magnitude.}
\label{plot8}
\end{figure}

The average differential extinction of stars as a function of $\upsilon_l$ and $\upsilon_b$
velocities is an important photometric quantity that can also be modeled and gives us information
about the dust within the NSC. Fig. \ref{plot8} shows this for the central field.  We notice that
the average differential extinction as a function of $\upsilon_l$ is negative for positive velocities
(preferentially at the front of the cluster) and positive for negative velocities (back of the
cluster) i.e., stars at the front of the cluster are observed with less extinction than their
neighbors. This finding is consistent with the asymmetry of the VHs in $l$ direction
(Fig. \ref{plot7}) and implies $A_K\simeq0.4$ within the NSC, see below. In contrast the average
differential extinction versus $\upsilon_b$ is relatively flat and consistent with the symmetric bell
shape of the VH in $b$ direction. However we still notice a scatter of the points which is
indicative of the systematic variations we should expect in $A_K$.

\begin{figure}
\centering
\includegraphics[width=\linewidth]{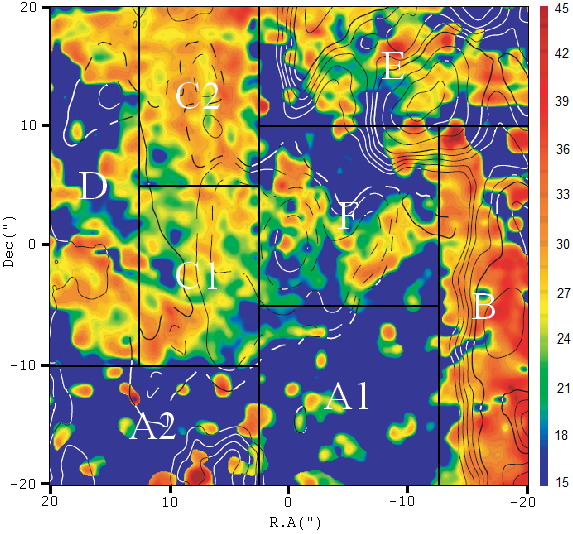}
\caption{Extinction towards the ionized gas in the galactic center, as derived in
  \citet[][their Fig.~5]{ss2003} from the ratio of their measured $\textrm{Pa}\alpha$ emission to
  the $\textrm{H} 92\alpha$ radio recombination-line emission of \citet{r1993}. The colours show
  extinction in the V-band according to the scale shown; dark blue signifies no information because
  of insufficient $\textrm{H} 92\alpha$ flux.  The field is split into eight cells, taking into
  account the outline of the mini-spiral, the shape of the $\textrm{VH}$s in $\upsilon_l$ and the
  $\delta A_K$ curves.}
\label{miniCellPlot}
\end{figure}

  Hot dust in the central parsec has been observed with mid-infrared emission
  \citep[e.g.,][]{lau2013}. The emission traces dust at the inner edge of the circumnuclear disk
  (CND) and in the ionized streamers collectively called the minispiral \citep[see
  Fig.~3,][]{lau2013}.  Fig.~\ref{miniCellPlot} shows a variable extinction distribution in our
  central field which overlaps this region. This figure was constructed by \cite{ss2003} using the
  ratio of their $\textrm{Pa}\alpha$ emission measurements to the $\textrm{H} 92\alpha$ radio
  recombination-line emission observed by \citet{r1993}.  The coloring signifies {\sl total}
  extinction including foreground along the line-of-sight and the contours show the outline of the
  mini-spiral.  Dark blue colour indicates insufficient signal-to-noise in the $\textrm{H} 92\alpha$
  emission.  The outline of the mini-spiral can be seen in this extinction map, and the variability
  of the dust extinction along the minispiral suggests that a fraction of this extinction is likely
  to be associated with the mini-spiral itself and therefore is located within the NSC. The outline
  of the minispiral in this map and in the MIR emission is closely similar, but the ratio of
  emission to extinction is not constant. In some regions outside the minispiral, particularly at
  larger radii, dust extinction is inferred from the ratio of $\textrm{Pa}\alpha$ to $6\textrm{cm}$
  radio continuum emission \citep{ss2003}, presumably arising at least partially from colder dust in
  the surrounding CND.  These considerations will motivate our choice of subfields for which we will
  construct separate dust models in Section \ref{sec:predictions}. For the development of the dust
  modeling in the next section we note that dust associated with the minispiral is likely to be
  concentrated in a small distance interval along the \los\, of order a fraction of the radius in
  the sky.

\section{DUST MODELING}

We saw in the previous section that the observed asymmetry of the $\textrm{VH}$ in $\upsilon_l$
is likely to be associated with dust extinction. In this section we describe how one can make an
analytical dust extinction model and use it with an already existing model of the NSC similar
to that of section 2.

For the rest of this work, along with $l$ and $b$ we use a Cartesian coordinate system $(x,y,z)$
where $z$ is parallel to the axis of rotation as before, $y$ is along the line of sight (smaller
values closer to the earth) and $x$ is along the direction of negative longitude, with the center
of the NSC located at the origin.

\begin{figure}
\centering
\includegraphics[width=\linewidth]{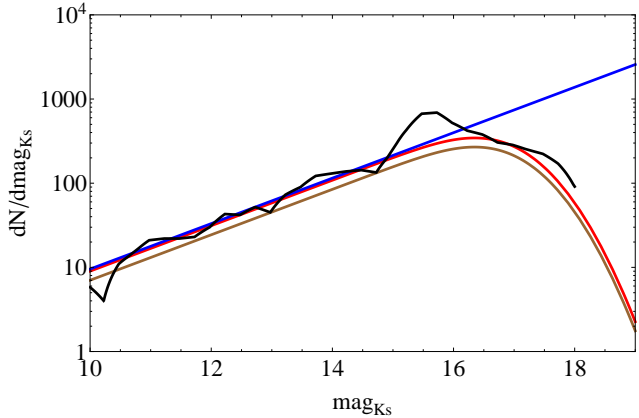}
\caption{Modeled K-band luminosity function without and with dust extinction. The blue line shows a
  power law fit to the stars $11 < K < 14$ with power-law index of $0.27\pm{0.02}$ as in
  \citet{schoedel2010}. The red line shows the effect of the completeness function. The brown curve
  is the red curve shifted faintwards by $A_K=0.4$ dust extinction.  The black line shows
  the K luminosity function for the Galactic center stars.}
\label{plot12}
\end{figure}

First we need to model a luminosity function. We can do that by taking the product
of two functions. The first represents a power law function in luminosity, corresponding
to an exponential magnitude distribution, $L(m)=10^{\gamma\cdot m}$. The second is an error function
that represents the completeness function, so that:

\begin{align}
\label{totalLum}
\begin{array}{l}
\frac{{dN}}{dm} = L(m) \times C(m) = \\
\\
{10^{\gamma\cdot m}}(1 - \textrm{erf}[(m - m_0)/\sigma])/2
\end{array}
\end{align}
Here $\gamma$ is the power law index of the luminosity function and $m_0$ is the
value where the completeness function $C(m)$ has its half height. For the power law we set the index
to $\gamma=0.27\pm{0.02}$ as in \citet{schoedel2010}. For the completeness function we set $m_0 =
16.5$ and $\sigma = 1$ because we found that these values represent well the K luminosity data as
shown in Figure \ref{plot12}. The red curve of Figure \ref{plot12} shows equation \ref{totalLum}
with the chosen values. We have also investigated models in which we approximated the red
clump bump at K$_S\simeq16$ in Fig.~\ref{plot12} by an additional Gaussian term. This led to
$\sim 10\%$ increase in the derived $A_K$ values. Since this is below the accuracy of our
determinations, we use the simpler power-law luminosity function model in the following.

\begin{figure}
\centering
\includegraphics[width=\linewidth]{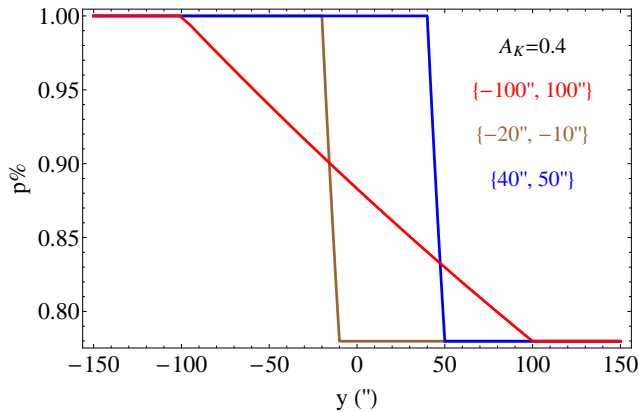}
\caption{Percentage reduction in observable stars after a dust effect with $A_K=0.4$.
as a function of \los\, for a homogeneous dust model (red) and two dust screen models (brown, blue).
About 75\% of stars remain in the observed luminosity function after the extinction due to dust.}
\label{plot6}
\end{figure}

Next we need the extinction variation over the \los\, which is just the derivative of the extinction
over the \los\, i.e. $d{a_K}/dy$. The function $d{a_K}/dy$ is general and could for example be represented
as a sum of Gaussians but for simplicity we choose a square function, so that:
\begin{align}
\label{extVar}
\frac{d{a_K}}{dy} = \left\{ {\begin{array}{l}
{0, y < {y_1}}\\
  c, {y_1 \leq y \leq y_2}\\
{0, y > {y_2}}
\end{array}} \right.
\end{align}
in which $y_1$ and $y_2$ indicate the positions where the dust starts and ends respectively.
The integral of eq.\ref{extVar} over all \los\, is the maximum extinction $A_{K}$
thus the constant c takes the value $c = {A_{K }}/\Delta y$ where $\Delta y=y_2-y_1$.
Function \ref{extVar} integrates to:
\begin{align}
\label{dAk}
{a_K}(y) = \left\{ \begin{array}{l}
0\,\,,\,\,y < y_1\\
\frac{{A_{K}}}{\Delta y}\left( y - y_1 \right), {y_1 \leq y \leq y_2}\\
{A_{K}}\,,\,\,y > y_2
\end{array} \right.
\end{align}
The percentage reduction in observed stars as a function of \los\, distance is:
\begin{align}
\label{p}
{p(y)} = \frac{{\int {L(m - {a_{K}(y)})C(m)dm} }}{{\int {L(m)C(m)dm} }}
\end{align}
With this we can calculate the percentage reduction in numbers of stars after extinction, 
${p_{\max }} = p(y_2)$. The percentage of stars hidden by extinction for $A_{K}=0.4$ is about 25\%.
For the simple case where the luminosity function is a power law, the previous
equation (\ref{p}) takes the form $p(y) = L(-{a_K}(y))$. The function for $A_{K}=0.4$ is shown
in figure \ref{plot6}. 

\begin{figure}
\centering
\includegraphics[width=\linewidth]{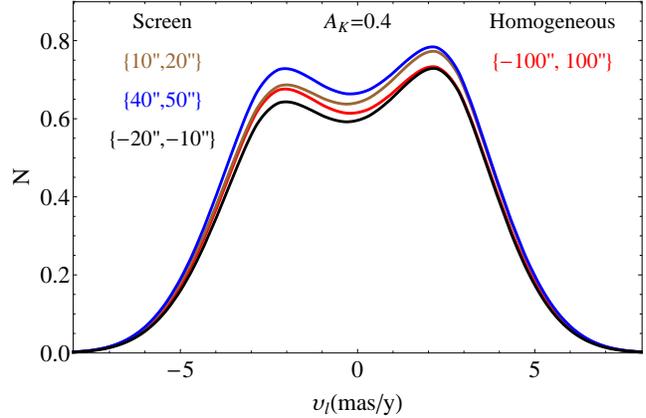}
\caption{VPDs in $l$ direction for different dust distribution models, as indicated on the plot. The
  right peak is higher due to the combined effects of dust extinction and rotation. We can achieve
  the same effect (e.g. the same asymmetry) for several combinations of $A_K$s and the distances
  where the dust is placed along the \los.  All VPDs are for a \los\, with $l=10''$ and
    $b=10''$.}
\label{plotE}
\end{figure}

Having that we calculate the $\textrm{VP}_l$ including the effect of dust extinction, denoted by $\textrm{VPD}$:
\begin{align}
\textrm{VPD}(\upsilon_{\rm l};x,z) = \frac{1}{\Sigma }\iiint\limits_{E>0} {p(y)f_{tot}(E,{L_z})\,d{\upsilon_{los}}d{\upsilon_z}dy}.
\label{VPldust}
\end{align}
Here $\Sigma$ is the stellar surface density of the NSC model, and
$f_{tot}(E,{L_z})={f_e}(E,{L_z}) + {f_o}(E,{L_z})$ is the total DF, consisting of an even part in
$L_z$ (contributing to the density) and an odd part (contributing to rotation), as in
\cite{cf2015}. Figure \ref{plotE} shows typical $\textrm{VPDs}$ after adding dust with
$A_{K}=0.4$. We observe that the right peak of the $\textrm{VP}_l$ is now higher than the left peak
which is a combined effect of dust and rotation. The dust does not produce an asymmetry for the VP
in the $b$ direction. One useful quantity is $A_K (\upsilon _l)$ averaged over the \los. This can be
calculated with
\begin{align}
\left\langle A_{K}(\upsilon _l;x,z) \right\rangle  = \frac{{\iiint {a_{K}(y)p(y) f_{tot} \left( E,L_z \right)d\upsilon _zd\upsilon _{los}}dy }} {{ \Sigma \times \textrm{VPD}\left( \upsilon _l \right) }}
\label{averageAk}
\end{align}
for each \los\, and connects our model with the photometry. From this we calculate the average
differential extinction between stars of given $\upsilon_l$ and their neighbors, which corresponds to the
data of Fig.~\ref{plot8}:
\begin{align}
\delta A_{K}(\upsilon _l;x,z) = \left\langle A_{K}(\upsilon _l) \right\rangle  - \frac{{\int {\left\langle {A_{K}\left( \upsilon _l \right)} \right\rangle \textrm{VPD}\left( \upsilon _l \right)d{\upsilon _l}} }} {{\Sigma\int {\textrm{VPD}\left( \upsilon _l \right)d\upsilon _l} }}
\label{averagedAk}
\end{align}

In order to understand the effects of dust extinction on $\delta A_{K}(\upsilon _l)$, we use two simple models for the dust distribution. The first is a homogeneous dust model that extends a few parsecs along the \los. The second is a thin $\sim10''$ screen of dust placed at several positions along the \los. We have verified that the width of the screen is not a sensitive parameter and the results are almost unchanged if we set it for example to $\sim5''$. Figures \ref{plotE} and \ref{plotABCD} shows these effects. In Fig. \ref{plotE} we see the effect of three dust screen models and one homogeneous model on the $\textrm{VP}_l$ for $A_K=0.4$. 

One important point to notice is that we can achieve the same effect (e.g. the same amount of
asymmetry in the histograms) with several models by using different combinations of $A_{K}$ and the
distances at which the dust is placed along the \los\,, therefore the dust extinction model is
degenerate. In Fig. \ref{plotABCD} we see plots of several screen and homogeneous models based on
eq. \ref{averagedAk}. The top panels show the $\delta A_{K}(\upsilon _l)$ curves of a screen dust
model placed at several distances in front of (left) and at the back (right) of the cluster for
$A_K=0.4$. The first thing to notice is that the further the screen of dust is placed from the
center the smaller is the effect of dust. This makes sense since far from the center the density of
stars is lower. We also note that if the screen of dust is placed in front of the cluster, the
curves are close to constant for the stars behind the cluster (i.e., for negative
$\upsilon_l$) since there is no dust there to affect the $\delta A_{K}(\upsilon _l)$. The
opposite happens when the screen of dust is behind the cluster.  The bottom left panel shows the
shape of $\delta A_{K}(\upsilon _l)$ for different $A_K$. The bottom right panel shows three
homogeneous models that extend over different distance intervals along the \los. In this case the
curves are symmetric relative to zero. The dust extinction model was implemented with \cite{w2011}.

\section{Predictions of the Dust model}
\label{sec:predictions}

In the last section we described how one can include the effects of
dust extinction in the dynamical modeling of the NSC, and calculated
differential reddening signatures for the NSC stars. Here we proceed
to model predictions and compare with both photometric and kinematic
evidence. We will see that the asymmetries seen in the VPs for $\upsilon_l$
can be explained as due to dust within the NSC. Our goal is to see
whether dust in the Galactic Center mini-spiral can explain our data,
and also to provide a rough extinction map of the central field based
on the model and the available data.

As discussed in Section~\ref{sec:NSCextinction}, MIR observations \citep[see the MIR map
of][]{lau2013} show that hot dust is associated with the minispiral within the NSC although
it cannot be assumed to trace the dust extinction directly.
Figure~\ref{miniCellPlot} based on Figure 5 of \citet{ss2003} shows the total dust extinction in the
central $40''\times40''$ around Sgr A*, derived from the observed ratio of $\textrm{Pa}\alpha$ to
$\textrm{H} 92\alpha$ radio recombination-line emission. This map includes both foreground and NSC
extinction. It shows the outline of the minispiral because it is limited by the signal-to-noise
ratio (S/N) of the $\textrm{H} 92\alpha$ recombination line flux, and the regions in which the
extinction has an apparent value of $A_V = 15$ is where the S/N is not sufficient to derive the
extinction. Fig.~\ref{miniCellPlot} shows the greatest extinction in front of the Western arm, and
relatively low extinction in front of the east-west bar. This suggests, in addition to possible
foreground extinction, position-dependent extinction associated with the minispiral itself.

For our modelling, Fig. \ref{miniCellPlot} is split into eight cells and sub-cells. The reasoning
behind the choice of these cells is based on the outline of the mini-spiral, but also on the
statistics and shape of the velocity histograms in $\upsilon_l$ and the mean differential extinction
variation $\delta A_K$ along \los\ in these cells. Some cells cover the region outside the
minispiral where the observed ratio of $\textrm{Pa}\alpha$ to 6 cm radio continuum emission
\citep{ss2003} also shows extinction, possibly in the foreground. This will also be tested by our
modelling. The $\textrm{VH}_l$s for all cells and the $\delta A_K$ as a function of $\upsilon_l$
and $\upsilon_b$ are shown in Fig. \ref{predictions} for the following cells:

\begin{itemize}
\item Cell A1: This cell's small $\delta A_K$ values are consistent with the lack of features in the extinction
map in comparison with other cells. Also the $\delta A_K$ values are consistent with the $\textrm{VH}_l$ since
both peaks look symmetric which is a sign of lack (or small amount) of dust within the NSC.

\item Cell A2: The cell lacks strong mini-spiral features as cell A1. However both the $\textrm{VH}_l$ and
$\delta A_K$ values show stronger effects of dust therefore this cell is separated from cell A1.

\item Cell B: The $\delta A_K$ values and the $\textrm{VH}_l$ are consistent with the strong features
of the mini-spiral in the extinction map since the $\delta A_K$ points are higher and lower for negative
and positive velocities respectively than the other cells and the asymmetry of the $\textrm{VH}_l$ is
intense. We also note that the dust effects are similar within the whole area B because after splitting
it into 2 sub-areas (not shown) we observed the same signature.

\item Cells C1 \& C2: The cells C1 and C2 belong to the Northern Arm of the mini-spiral. The shape
of the $\delta A_K$ data for both cells seems similar for negative velocities but the $\delta A_K$
for C2 is more symmetric and consistent with the extinction map hence we split the area into 2
halves. We notice also some asymmetry on the $\textrm{VH}_l$s of both areas C1 and C2.

\item Cell D was separated from C1 \& C2 because the $\delta A_K$ values look more symmetric than
C1 \& C2.  We also note that the dust effects are similar within the whole area D because after splitting
it into 2 sub-areas (not shown) we observed the same signature.

\item Cells E \& F: The  $\delta A_K$ values of these two cells look similar but the $\textrm{VH}_l$ of the cell F
lacks the asymmetry characteristic in contrast of cell E therefore we keep them separate. 
\end{itemize}

The central field with these eight cells is surrounded by a a more
extended area with observations for about 2000 stars. We split this
area into three outer fields O1-O3 placed around the central field as
shown in Fig. \ref{plot2}. The $\textrm{VH}_l$ for these cells and the $\delta A_K$ as a function of
$\upsilon_l$ are shown in Fig. \ref{outer}.

\begin{figure*}
\centering
\includegraphics[width=\linewidth]{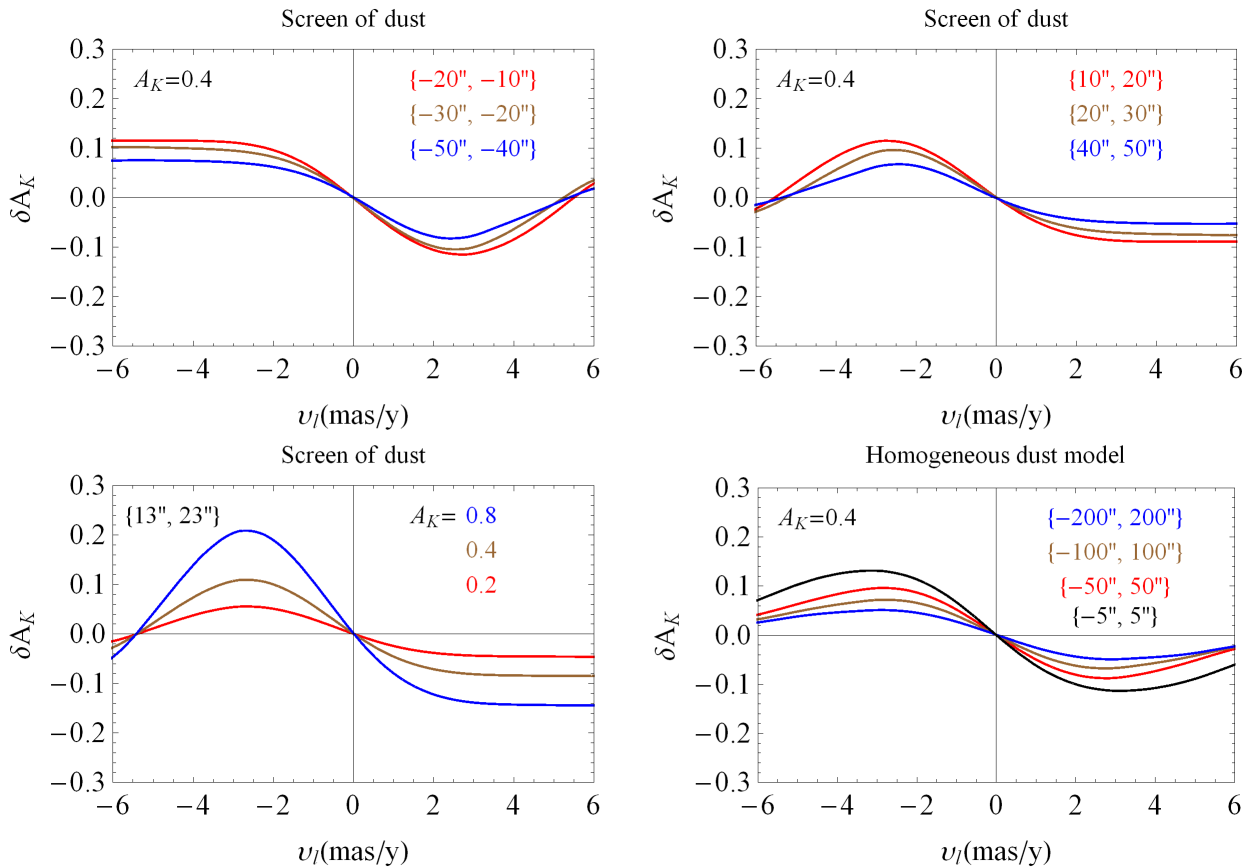}
\caption{Shape of $\delta A_{K}$ versus $\upsilon_l$ curve for several dust models. Top left: in
  front of the cluster for constant $A_K=0.4$.  The numbers in the parenthesis show where the dust
  screen starts and ends respectively along the \los.  Top right: Dust screen models placed behind
  the cluster for constant $A_K=0.4$.  Bottom left: Dust screen models placed slightly behind the
  cluster for several $A_K$ values.  Bottom right: Homogeneous dust models that extend over
  different distances ranges along the \los\, for constant $A_K$. The \los\, for all curves has
  coordinates $l=10''$ and $b=10''$.}
\label{plotABCD}
\end{figure*}

\begin{figure*}
\centering
\includegraphics[scale=0.41]{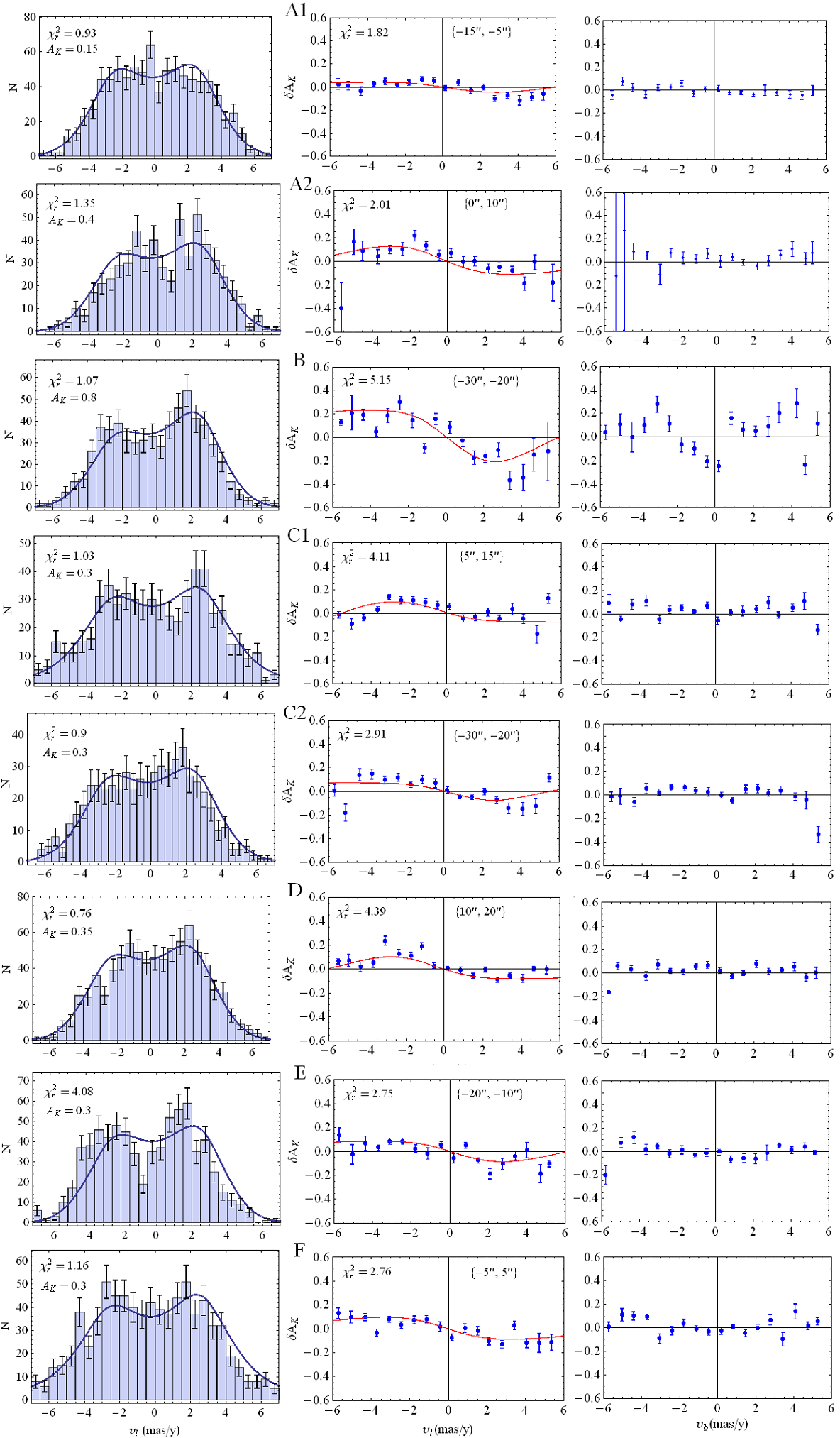}
\caption{Predictions of the model with the VH and $\delta A_{K}$ data for each cell. The numbers in the brackets
show where the screen of dust is placed relative to the center. The \los\, is placed at the center of each cell. Reduced $\chi^2$ are also provided for the histograms and the photometry. For the cells A1, A2, B, C2, E, the screen dust distance is based on the orbit models of the mini-spiral \citep{zm2009}.}
\label{predictions}
\end{figure*}

\begin{figure*}
\centering
\includegraphics[scale=0.7]{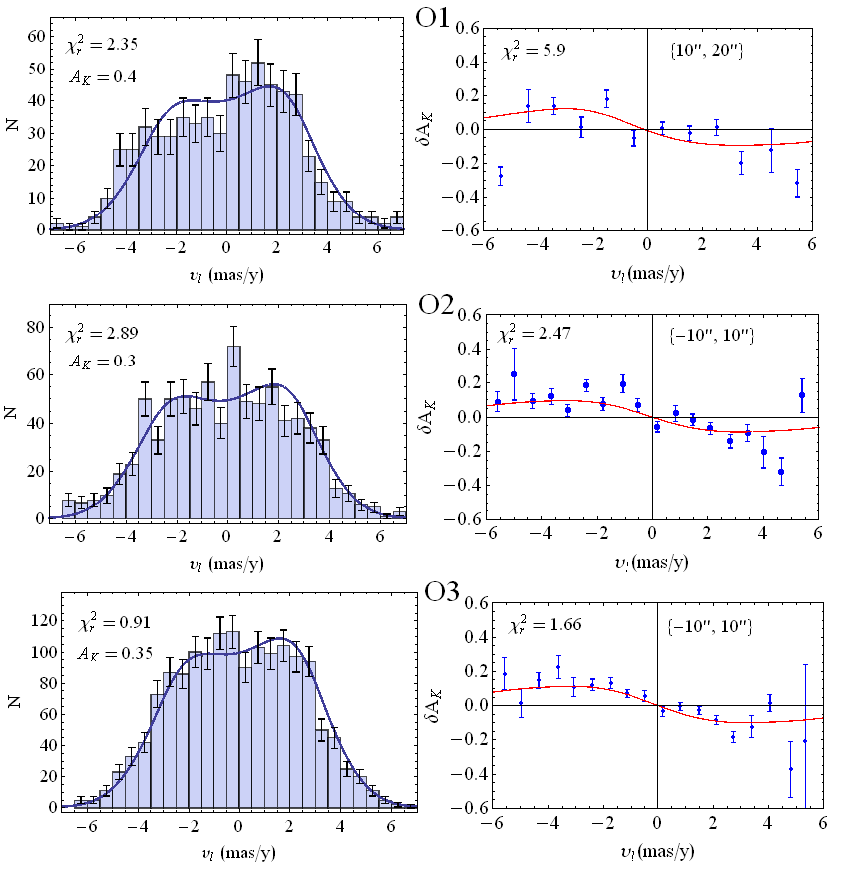}
\caption{Predictions of the model with the VH and $\delta A_{K}$ data for the outer cells shown in Fig.\ref{plot2}. The numbers in the brackets show where the screen of dust is placed relative to the center. The \los\, is placed at the center of each cell. Reduced $\chi^2$ are also provided.}
\label{outer}
\end{figure*}

Our goal is to give a model prediction of each of these cells (8+3 in
total). For the model we use a thin screen of dust with width $10''$
because the dust associated with the minispiral is likely to be
concentrated in a small distance interval along the \los. The precise
width of the dust screen is not important as the dust signatures are
insensitive to this parameter. The two main parameters are the total
extinction in the screen and its location along the \los. However, as
explained in the last section, these two parameters are partially
degenerate. The degeneracy is particularly strong for the $\textrm{VH}_l$
histograms.  The $\delta A_{K}$ data in principle are sensitive to whether the
extinction is in front or behind the Galactic center, as shown in Fig. \ref{plotABCD}.
However, the $A_{K}$ data have large scatter between adjacent data points such that points
with seemingly small error bars can even have the `wrong' sign of $\delta A_{K}$
(Fig. \ref{predictions}). This large scatter is also seen in the $\delta A_{K}$ versus $\upsilon_b$
plots (also shown in Fig. \ref{predictions}) where no dust signature is present. Therefore we
decided to not try to fit the data using $\chi^2$.

Rather, we choose to place the dust screens along the \los\, according to
other available information, and only deviate from this when this
appears inconsistent with the shape of the $\delta A_{K}$ distribution.  The
total dust extinction of the dust screen is then chosen by eye mostly
from the amplitude and shape of the $\delta A_{K}$ distribution, taking into
account also the scatter of the $\delta A_{K}$ points, and to a lesser degree
from the asymmetry of the $\textrm{VH}_l$ peaks.

Specifically for the central field we use the three orbit-model of
\cite{zm2009} for the three ionized gas structures in the central
3pc (the Northern Arm, Eastern Arm, and Western Arc). We then map
the center of each of the cells to a point in the relevant orbit
plane according to its R.A and Dec. position. The distance from the
center along the \los\, is given from the coordinates of that point on
the orbit plane. For each cell, we use one common mean distance. Table
\ref{table1} shows to which orbital plane each cell is assigned, and the
distance of the dust screen from Sgr A*. 

Fig. \ref{predictions} shows the predictions of the screen dust model for the 8 cells of the central field.
The reasoning behind the choice of $A_{K}$ and positions for the dust screen is based on the shape of the data and
the examples of Fig. \ref{plotABCD}:

\begin{itemize}
\item Cell A1 \& B: We place the dust screen in front of the cluster according to the value of Table \ref{table1}.
These two cells are interesting because they both show a correlation between the outline of the mini-spiral and
the photometry data. They also exhibit the maximum contrast between the amount of dust. Cell B needs $\sim5$ times more extinction than cell A.

\item Cell A2: The dust screen is placed behind the center according to Table \ref{table1}.

\item Cell C1 \& D: The shape of the data in conjunction with the top right panel of Fig. \ref{plotABCD}
indicate that the dust screen should be behind the cluster (also gives a much better $\chi^2$)
in contrast with the value of Table \ref{table1}. A possible reason for this could be that
the single orbit description is not accurate for the centers of these cells.

\item Cell C2 \& E: For both cells the dust screen is placed in front of the cluster according to
Table \ref{table1}.

\item Cell F: The shape of the data in conjunction with the lower right panel of Fig. \ref{plotABCD} indicate that
the dust screen should be centered.
\end{itemize}

For the outer cells of the extended field the $A_{K}$ is selected according to the general characteristics
of plot \ref{plotABCD}:
\begin{itemize}
\item Cell O1: the VHs and the photometry have a consistent signature. The extinction is close to
  $A_{K}\simeq0.4$ mag.
\item Cells O2 and O3: For these cells the photometry gives $A_{K}\simeq0.3$ and $0.35$ but the
  peaks of the VHs are almost symmetric. 
\end{itemize}
Based on the model of the CND in \citet{lau2013}, we might have expected stronger extinction effects in
fields O1 and O3 than in O2. However, this is not confirmed by Fig.~\ref{plotABCD} and the lack of correspondence
between the $\delta A_{K}$ and the VHs is puzzling.

Having the prediction for the cells we can estimate the foreground extinction for each cell using the $A_{K}$ values of Fig. \ref{plot2}. The second row of table \ref{table2} shows the total extinction of each cell of the central field based on Fig. \ref{plot2}. The third row shows an estimate of the foreground extinction based on $A_{K}\textrm{(total)} = A_{K}\textrm{(foreground)}+ x * A_{K}\textrm{(NSC)}$ where we approximate $x=0.5$ to take into account that of order half of the stars are unobscured because they are in front of the cluster. The exact value is not important for the scope of this paper but it should be closer to $x=0.5$ than
$x=1$.

\begin{table}
\begin{tabular}{llrll}
\hline\hline
Northern Arm & E (-15'') &             &             &     \\
Eastern Arm  & A2 (5'')  & C1 (-9'')& C2 (-25'')& D (-20'')\\
Western Arc  & B (-25'') & A1 (-10'')& F (-5'')  &			\\
\hline\hline
\end{tabular}
\caption{Each one of the eight cells of Fig. \ref{plotABCD} belongs to an orbital plane \citep{zm2009} representing one of the three ionized gas (Northern Arm, Eastern Arm, and Western Arc) formations. The inferred \los\, distance of the  dust screen from Sgr A* is given in the parentheses (negative points towards the earth).}
\label{table1}
\end{table}

\begin{table*}
\begin{tabular}{llrllllll}
\hline\hline
Cell                  & A1   & A2   & B   & C1   & C2   & D    & E    & F    \\
Total extinction      & 2.73 & 2.99 & 3.1 & 2.82 & 2.87 & 2.78 & 2.84 & 2.95 \\
Foreground extinction & 2.66 & 2.79 & 2.7 & 2.67 & 2.72 & 2.61 & 2.69 & 2.8 \\
\hline\hline
\end{tabular}
\caption{Extinction values per cell based on Fig.~\ref{plot2}. The foreground extinction of each cell is estimated 
according to $A_{K}\textrm{(total)} = A_{K}\textrm{(foreground)}+ x*A_{K}\textrm{(NSC)}$ where we approximate 
$x=0.5$ to take into account that of order half of the stars are unobscured because in front of the 
cluster. The exact value is not important for the scope of this paper but it should be closer to $x=0.5$ than $x=1$.}
\label{table2}
\end{table*}

\section{Does the addition of dust affect the measured $M_{\bullet}$, $M_*$ and $R_0$?}

In this section we try to answer how much the dust \emph{within} the NSC will affect the derived \citep{cf2015} statistical parallax, supermassive black hole and stellar mass of the NSC. The \emph{foreground} dust will affect the VPs only by a scale factor which does not impact the derived values. In \cite{cf2015} we derived new constraints on the $R_0$, $M_*$ and $M_{\bullet}$ by fitting to the corresponding data the $\left\langle{{\upsilon ^2}}\right \rangle_{l,b,los}^{1/2}$ parts of the $2^{nd}$ order Jeans moments, that are moments of the even part of the corresponding VPs of the 2-Integral distribution function. Therefore here the problem is reduced to how much the even part in $L_z$ of the VPs changes after the addition of dust within the NSC. 

\begin{figure}
\centering
\includegraphics[width=\linewidth]{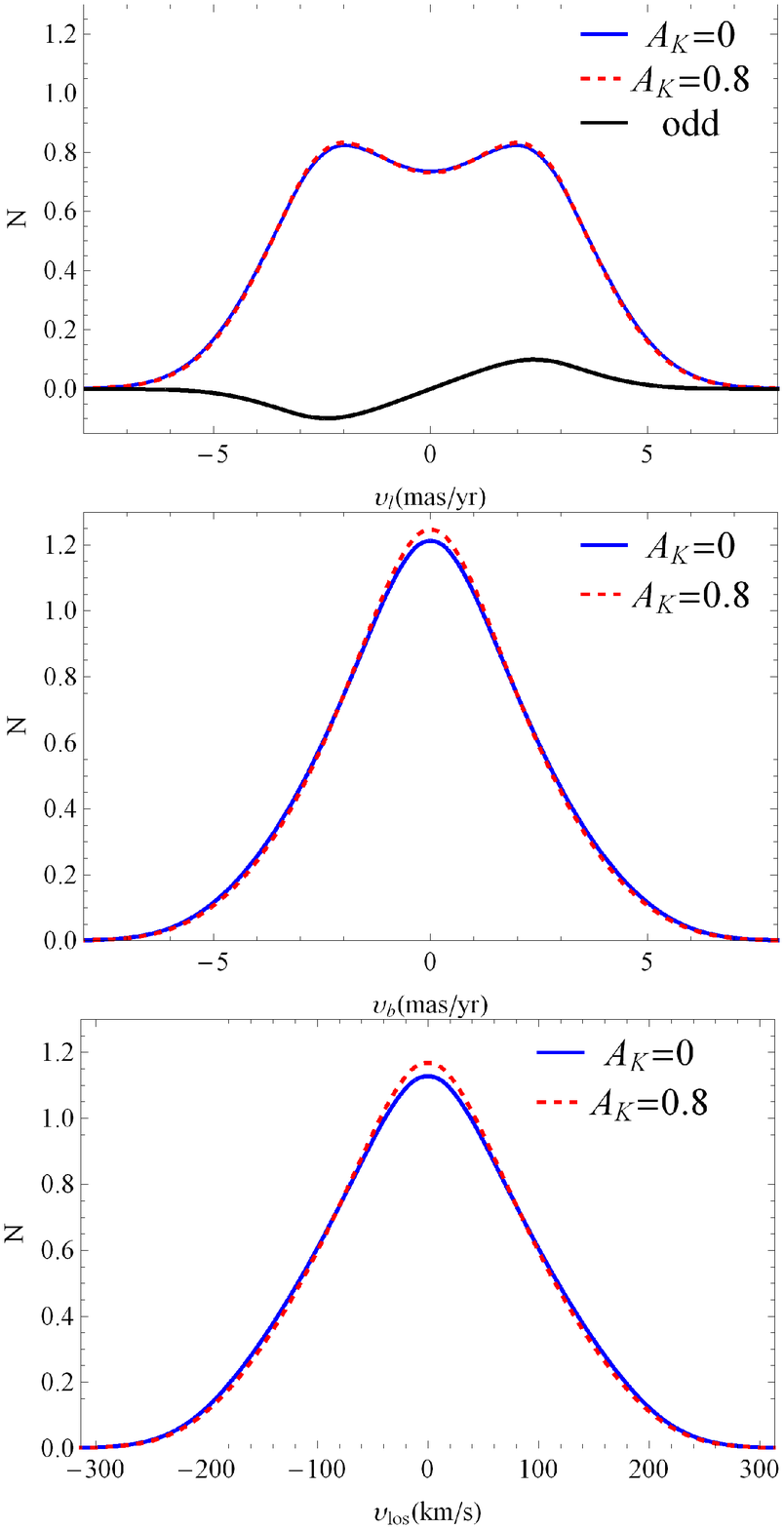}
\caption{Even part of the renormalized VPs for $\upsilon_l, \upsilon_b$ and $\upsilon_{los}$ for the NSC
dynamical model with no dust from \citet{cf2015}, and for the same dynamical model including the dust
screen prediction based on cell B. The black line in the upper panel corresponds to the odd part of the
$\textrm{VP}_l$. The VPs are calculated at the center of cell B.}
\label{VPDustNoDust}
\end{figure}

Figure \ref{VPDustNoDust} shows the even parts of the VPs for $\upsilon_l, \upsilon_b$ and
$\upsilon_{los}$ for the NSC dynamical model from \cite{cf2015} with best mass and distance
parameters, and for the same dynamical model including the screen dust prediction, for a \los\,
through cell B which has the largest amount of extinction ($A_K=0.8$) among the cells of the central
and outer fields. We notice that the difference between the VPs for this amount of extinction is
very small. Specifically the average difference of the $2^{nd}$ moments of the VPs between the two
models is $\sim1.5\%$. If instead we use $A_K=0.4$, close to the average extinction within the NSC
inferred from this work, the difference is smaller than $0.5\%$. The relative differences of
$\sigma_{los}/\sigma_b$ and $\sigma_{los}/\sigma_l$ between the model with no dust and the model
with $A_K=0.4$ are similarly small, $0.2\%$ and $0.6\%$, respectively. Therefore we conclude that
the systematic effects on the statistical parallax due to dust are within the estimated errors of
\cite{cf2015}, causing the distance to the NSC to decrease by $\sim 0.4\%\simeq 30$pc.

That the changes in the even part are so small can be explained by the following formal argument for the
$\textrm{VP}_l$. We show that for small amounts ($1^{st}$ assumption) of homogenized ($2^{nd}$ assumption)
dust around the center the even part of the renormalized $\textrm{VP}$ in $\upsilon_l$ is the same
as that for no dust. The odd part is a direct indicator of dust at that point $(l, b)$ and can be used to
estimate $A_{K}$.
\newline

When the luminosity function is a power law and the dust is homogeneously placed i.e. $y_1=-y_2$ in
equations \ref{extVar} and \ref{dAk}, $p(y)$ takes the form within the dust area:

\begin{align}
p(y)=L(-{a_K}(y))={10^{ -\gamma\left( {\frac{{A_{K }}}{2} + \frac{A_{K } y }{\Delta y}} \right)}}
\end{align}
For small $A_{K} \ll (\gamma\ln(10))^{-1}\simeq 1.61$ we have:

\begin{align}
\begin{array}{l}
p(y) \simeq 1 - \gamma\ln(10)\frac{{{A_{K}}}}{2} - \gamma\ln(10)\frac{{{A_{K}}}}{{\Delta y}}y\\
 = p(0) - ky
\end{array}
\end{align}
where $p(0)=1-\gamma\ln(10)\frac{{{A_{K}}}}{2}$ and $k=\gamma\ln(10)\frac{{{A_{K}}}}{{\Delta y}}$.

Therefore\footnote{we showed this for the case where the luminosity function is a power law
but the same holds in the general case $L(m)$ where $\gamma\cdot \ln(10)$ is replaced by $\frac{{\int {L'(m)C(m)} dm}}{{\int {L(m)C(m)} dm}}$.} the function $g(y)=p(y)-p(0)$ is odd everywhere (including the area where there is no dust).
Next for simplicity we use ${f_{tot}}(E,{L_z}) = f(E,x\upsilon_{los}-y\upsilon_l) \to f({\upsilon _l}y)$
because $y$ appears within $f$ only with the form of $y^2$ and $\upsilon_l y$.

Now we have:
\begin{align}
\begin{array}{l}
\frac{1}{2}\int {p(y)\iint {\left( {f({\upsilon _l}y) + f( - {\upsilon _l}y)} \right)d{\upsilon _z}d{\upsilon _{los}}dy} } \\ \\
 - \frac{1}{2}\int {p(0)\iint {\left( {f({\upsilon _l}y) + f( - {\upsilon _l}y)} \right)d{\upsilon _z}d{\upsilon _{los}}} dy} = \\ \\
 \frac{1}{2}\int {g(y) \iint {\left( {f({\upsilon _l}y) + f( - {\upsilon _l}y)} \right)d{\upsilon _z}d{\upsilon _{los}}} dy}  = 0
\end{array}
\end{align}
The previous is 0 because $g(y)$ is an odd function and $\iint {\left( {f({\upsilon _l}y)
+ f( - {\upsilon _l}y)} \right)d{\upsilon _z}d{\upsilon _{los}}}$ is an even function of $y$
therefore:

\begin{align}
\begin{array}{l}
\frac{1}{2}\int {p(y)\iint {\left( {f({\upsilon _l}y) + f( - {\upsilon _l}y)} \right)d{\upsilon _z}d{\upsilon _{los}}} dy} =\\ \\
\frac{1}{2}p(0)\iiint {\left( {f({\upsilon _l}y) + f( - {\upsilon _l}y)} \right)d{\upsilon _z}d{\upsilon _{los}}dy} 
\end{array}
\end{align}
And thus:
\begin{align}
\textrm{VPD}_{\textrm{even}}({\upsilon _l}) = p(0)\textrm{VP}_\textrm{l}({\upsilon _l})
\end{align}

To find the constant $p(0)$ we integrate once more over the velocity this time:
\begin{align}
\int { \textrm{VPD}_{\textrm{even}}  ({\upsilon _l})d{\upsilon _l}}  = p(0)\int {\textrm{VP}_\textrm{l}({\upsilon _l})d{\upsilon _l}}  = p(0)
\end{align}
Therefore $p(0)$ is the normalization factor of the $\textrm{VPD}_{\textrm{l}}$.
The above means that to first order, dust does not affect the even part of the VPD in $l$ direction significantly
except for a scale factor. Fig \ref{VPDustNoDust} shows that the effects for $\textrm{VPD}_{\textrm{b}}$ and $\textrm{VPD}_{\textrm{los}}$ are similarly small. If extinction within a stellar system is small
enough ($A_{K}\ll 1.6$ for the NSC) then fitting the even part of a model's VPs to the even part of the VHs
is sufficient to get accurate estimates of the $M_{\bullet}$, $M_*$ and $R_0$ parameters. In \cite{cf2015}
we used the root mean square velocities that are moments of the even parts of the VHs to fit the $M_{\bullet}$,
$M_*$ and $R_0$ parameters of the axisymmetric model. Therefore we expect that their values will not be
affected by dust more than $0.4\%$ as we explained earlier.

In principle the odd part of the $\textrm{VPD}_{\textrm{l}}$ can be fitted to the odd part of the VHs
and this part is scale free since the scaling factor is already known from the even part therefore one
can fit the $A_{K}$ for some combination of cells.

\section{Discussion \& Conclusions}
\label{discussion}

The main goal of this work was to understand the slight asymmetries in the $\textrm{VP}_l$s
of the NSC and their influence on the dynamical modeling following the recent work of \cite{cf2015}.
Our interest was triggered by the observation that the right peak of the $\textrm{VH}_l$ is
often slightly higher than the left. A plausible explanation was given based on the existence
of dust \emph{within} the NSC. Because of the dust, the apparent number of stars behind the NSC
is smaller than that in front of the cluster. This in conjunction with the rotation can explain the
observed characteristic.

In order to quantify the dust effects, we worked with proper motions and photometry for $\sim7100$ stars from
\cite{fc2014}. We applied an analytic dust extinction model together with our recent NSC dynamical model.
The extinction model gave us reasonable results and was able to predict both the signature in the
VPs and the photometry.

Observation of the NSC in the optical is almost impossible because of $~30$ mag extinction.
In the infrared the situation is much better since $A_{K}\sim 3$ mag. Most of this extinction
belongs to the foreground and does not affect the shape of the VHs (except a normalization
factor). We find here that a small fraction of the total extinction value $(\sim15\%)$ belongs
within the NSC.

The area between $\sim1-1.5$pc radius consists of several streamers of dust, ionized and atomic gas
with temperatures between $100K-10^4K$ and is called "ionized central cavity" \citep{e1983}. The
mini-spiral is a feature of the ionized cavity, and is formed from several streamers of
gas and dust infalling from the inner part of the CND \citep{ke2012}. It consists of four main
components: the Northern arm, the Eastern arm, the Western arm and the Bar \citep{zm2009} that can
be described well by streams of ionized gas or filaments orbiting Sgr A* \citep{sl1985}.

We investigated how well the mini-spiral correlates with the extinction effects in the NSC data within the central field. To assess this we first investigated whether our extinction model puts the dust on the same side of the NSC as does the mini-spiral interpretation. This is true for six out of the eight cells (except cell C2). We found the largest extinction $(A_K=0.8)$ in cell B where also the largest extinction is inferred from the extinction map of \cite{ss2003}, and particularly low extinction in cell A where the extinction map is consistent with only foreground extinction.

We can estimate the mass that corresponds to a given amount of extinction using
\begin{align}
{\rho _d} = \frac{{{A_{K}}}}{{\Delta y}}\frac{1}{{1.086{\kappa _\lambda }}}
\end{align}
where ${\rho _d}$ is the dust density, ${{\kappa _\lambda }} = 1670\textrm{cm}^2/\textrm{g}$ \citep{d2003a} is the mass extinction coefficient \footnote{ftp://ftp.astro.princeton.edu/draine/dust/mix/kext\_
\\ \_albedo\_WD\_MW\_3.1\_60\_D03.all} for the K-Band, ${{A_{K}}}=0.4$ from the model
prediction and ${\Delta y}=10''$ is the width of the dust screen. We find that
$\rho _d \simeq 1.8\times{10^{-19}} \textrm{kg}/\textrm{m}^3$ and the dust mass within a parallelepiped with dimensions $(10'',40'',40'')$ centered on Sgr A*  is ${M_d} \sim 3{M_\odot}$. Since the dust extinction model presented here is not precise we consider that this estimate is correct only within an order of magnitude. This value is within the range of $0.25-4M_\odot$ for the mini-spiral found from other works \citep{zm1995,ke2012,es2011}.

Finally we showed that for small values of extinction the even parts of the VPs are not affected significantly. As a result, the measured $M_{\bullet}$ and $M_*$ parameters of \cite{cf2015} do not change by more than $\sim0.4\%$ for extinction $A_K\simeq0.4$ and the inferred $R_0$ is decreased by the same amount, which is less than the smallest systematic error (for the statistical parallax) inferred for these parameters in \cite{cf2015}.
\newline\newline
Our results can be summarized as follows:
\begin{itemize}
\item We showed that extinction due to dust explains kinematic asymmetries and differential photometry of the
NSC, and measured the amount of extinction within the NSC by combining a dynamical model with a dust extinction
model.
\item We presented an extinction table for the dust \emph{within} the NSC in several cells.
\item We found that the distribution of the dust is consistent with the extinction being associated with the
mini-spiral for six out of eight cells.
\item Systematic effects due to dust with typical extinction $A_K\simeq0.4$ affect the $M_{\bullet}$, $M_*$ and $R_0$ parameters deduced from previous dynamical modeling only by $\simeq0.4\%$, which is smaller than their estimated systematic errors.
\end{itemize}


\FloatBarrier

\label{lastpage}

\end{document}